\documentclass[twocolumn,tighten]{aastex63}
\usepackage{graphicx}
\usepackage{gensymb}
\usepackage{longtable}
\usepackage{amsmath}

\shorttitle{Nature of the IC 1613-S8 Supernova Remnant}
\shortauthors{Fesen and Weil}

\begin{document}

\title{The Nature of the Young Supernova Remnant S8 in the Dwarf Galaxy IC 1613 }

\author[0000-0003-3829-2056]{Robert A.\ Fesen}
\affil{6127 Wilder Lab, Department of Physics and Astronomy, Dartmouth
                 College, Hanover, NH 03755 USA}

\author[00000-0002-4471-9960]{Kathryn E.\ Weil}
\affil{6127 Wilder Lab, Department of Physics and Astronomy, Dartmouth
                 College, Hanover, NH 03755 USA}

\begin{abstract}

We present sub-arcsecond optical images and low- to moderate-resolution spectra
of the compact, X-ray and optically bright supernova remnant known as S8 in the
nearby dwarf galaxy IC 1613. Deep H$\alpha$ images of the remnant show a
sharply defined crescent shaped nebula, while narrow passband images reveal a
coincident and unexpectedly bright continuum nebulosity exhibiting a size and
morphology like that seen for the remnant's line emissions.  Low-dispersion
spectra covering $3600 - 9000$ \AA \ show numerous low-ionization line
emissions such as [\ion{O}{1}] and [\ion{Fe}{2}], along with higher-ionization
emission lines including \ion{He}{2} and optical coronal lines [\ion{Fe}{7}],
[\ion{Fe}{10}], [\ion{Fe}{11}], and [\ion{Fe}{14}]. This suggests the presence
of a wide range of shock velocities from $\sim$ 50 to over 350 km s$^{-1}$,
corresponding to preshock densities of $\sim1 - 30$ cm$^{-3}$.  Higher
resolution spectra indicate an expansion velocity around 180 km s$^{-1}$ with a
$\sim45$ km s$^{-1}$ wide central cavity. H$\alpha$ emission spans rest frame
velocities of $+120$ to $-240$ km s$^{-1}$ and we estimate a total nebula mass
of $119 \pm 34$ M$_{\odot}$. We conclude S8 is a relatively young supernova
remnant ($\simeq2700 - 4400$ yr) exhibiting properties remarkably like those
seen in the young LMC remnant N49, including age, physical size, shock
velocities, filament densities, optical line strengths, X-ray and optical
luminosities, and coronal line and continuum emissions.

\end{abstract}
\bigskip

\keywords{SN: individual objects: ISM: supernova remnant} 

\section{Introduction}

Studies of young core-collapse supernovae (CCSNe) remnants are useful for
understanding high mass stellar explosions, ejecta asymmetries, and
the formation of central compact objects \citep{Chevalier2005,Patnaude2017}.
Unfortunately, young CCSN remnants with ages less than about 5000 yr are
relatively rare as there are only about a dozen known in the Milky Way, and just 
a few more in the Magellanic Clouds and other extragalactic systems
\citep{Mili2017,Branch2017}. 

One of the few suspected young extragalactic supernova 
remnants (SNRs) is a
small compact remnant, S8, located in the outskirts of the Local Group dwarf
irregular galaxy IC 1613 (d = 725 kpc; \citealt{Hatt2017}).  Labeled by
\citet{Sandage1971} as emission nebula No.\ 8 in IC~1613, hence the name S8, it
was initially seen as just one of several H II regions in the northeastern
region of IC~1613.  However, it was subsequently recognized as a SNR based on
optical emission line ratios and nonthermal radio flux index
\citep{Smith1975,DDB1980,Dickel1985,Peimbert1988}. 

Classified as a small composite SNR consisting of a bright radio and optical
emission shell and an X-ray bright extended nebula \citep{Schlegel2019}, the remnant
appears in the optical as a bright crescent shaped emission nebula $3\farcs5
\times 5\farcs0$ in size (12 pc x 18 pc at 725 kpc) Its optical spectra have been  
modeled with shock velocities $\sim$50--150 km s$^{-1}$. It exhibits an
expansion velocity of 250--300 km s$^{-1}$, low metallicity reflective of IC
1613's abundances, and [\ion{S}{2}] line emissions indicating a relatively high
electron density of $\simeq 1400$ cm$^{-3}$, consistent with a young
remnant \citep{Lozinskaya1998,Lozinskaya2009}.

With an estimated age of  $\sim$ 3400--5600 yr, the S8 remnant is quite 
luminous in the optical and X-rays, with an X-ray luminosity of $5.6 \times 10^{36}$ erg s$^{-1}$
\citep{Schlegel2019} ranking it among
the brightest X-ray SNRs in the Local Group.  It is also bright in the radio, with a luminosity nearly 20\%
that of the Crab Nebula at 20 cm, with a spectral index of $-0.56 \pm
0.06$ consistent with a SNR's nonthermal emission associated with ISM shocks \citep{Lozinskaya1998}.
Given its location in among IC~1613's northeastern H~II regions 
and large superbubbles, formed via stellar winds
from massive stars, plus its size and X-ray luminosity S8 is a suspected CCSN remnant \citep{Rosado2001,OU2018,Schlegel2019}.

\begin{figure*}[ht]
\centerline{\includegraphics[angle=0,width=17.cm]{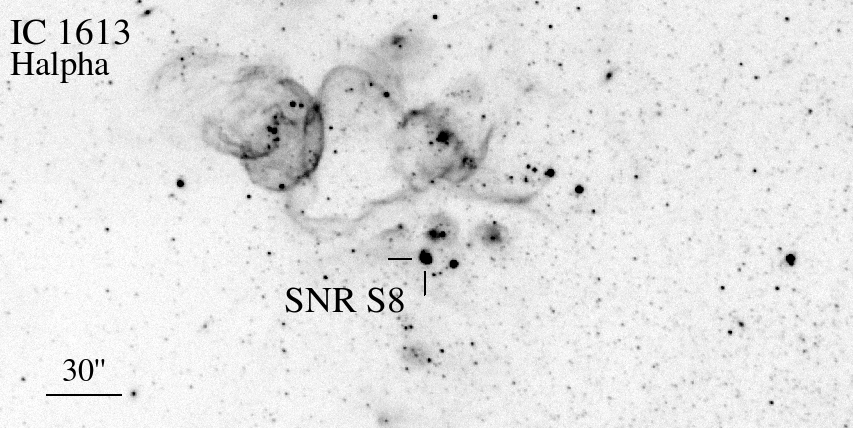}}
\caption{H$\alpha$ image of the H II regions in the northern area of IC~1613
around the location of the supernova remnant S8.
North is up, East is to the left. }
\label{WideFOV}
\end{figure*}

Due to its suspected youth, high luminosity and expansion velocity, S8 has
been compared to some young LMC core-collapse SNRs
\citep{Lozinskaya1998,Schlegel2019}.  Here we present sub-arcsecond
H$\alpha$ and continuum filter images along with low and moderate resolution
optical spectra of the S8 remnant obtained in order to explore its optical
properties and hence its nature in relation to other relatively young SNRs. Our
data are described and presented in $\S$2 and $\S$3. In $\S$4, we discuss this
remnant's general optical properties, possible origins of its unusually
bright and extended continuum emission, and compare its optical and X-ray
properties to young CCSN remnants in the LMC and nearby galaxies.

\bigskip

\section{Observations}

A series of both broad and narrow passband interference filter images of SNR S8 were
obtained in October 2019 and January 2020 with the 2.4m Hiltner telescope at
the MDM Observatory at Kitt Peak, Arizona using the Ohio State Multi-Object
Spectrograph (OSMOS; \citealt{Martini2011}).  Images were taken using a V band
filter, an [\ion{O}{3}] 5007 \AA \ filter (FWHM = 80 \AA), an H$\alpha$ filter
(FWHM = 80 \AA), and a broadband R filter matching that of the ACS F675W filter
aboard the {\sl Hubble Space Telescope } ({\sl HST}).  Additional images were
taken using narrow passband filters centered on spectral regions free of any of 
S8's strong line emissions based on previously published spectra
\citep{Peimbert1988,Lozinskaya1998}.  These continuum filter images included a blue
continuum filter ($\lambda_{\rm c}$ = 5288 \AA, FWHM = 256 \AA), and two red
continuum filters centered on either side of H$\alpha$; $\lambda_{\rm c}$ = 6071
\AA (FWHM = 260 \AA), and $\lambda_{\rm c}$ = 7021 \AA \ (FWHM = 177 \AA).
Multiple exposures were taken in each filter with exposures ranging from 30 s
for the F675W filter to 1200 s for the narrow passband filters. 
Seeing was very good for most images, with measured 
FWHM values for individual 7021 \AA \ filter images ranging between 
$0\farcs80 - 0\farcs84$ and $0\farcs90 - 0\farcs95$ for H$\alpha$ images. 

We note that the 6071 \AA \ continuum filter's transmission window covers an
almost completely emission line free portion of the remnant's optical spectrum
(see \S3.2 below). The filter's transmission curve has sharp cut-on/cut-offs at
5900 and 6200 \AA,  meaning that with the exception of  weak [\ion{Fe}{7}] 6087
\AA \ emission, the filter provides a $\sim$300 \AA \ wide emission line free
bandpass.  Similarly, the 7021 \AA \ filter's bandpass of $\sim$250 \AA (6900
to 7150 \AA) is sensitive to only \ion{He}{1} 7065, [\ion{Ar}{3}], and
[\ion{Fe}{2}] 7155 which are relatively weak in S8 (see \S3.2 below).

Images taken in October 2019 were followed by a series of low dispersion,
long-slit OSMOS spectra using two different grism setups.  Using a 1.4 arcsec
N-S slit and a red VPH grism (R = 1600), we obtained $2 \times 2000$ s spectra
covering $4500 - 8500$ \AA \ with a measured FWHM $\simeq$6.5 \AA. This was followed
by two blue VPH grism set-ups using a 1.2 arcsec slit and covering the
wavelength regions $3600 - 5900$ and $4000 - 7000$ \AA \ (FWHM = 3.5 \AA).
Exposure times ranged from 1200 to 3000 seconds. The resulting spectra were wavelength
corrected by $-230$ km s$^{-1}$ based on the measured rest velocity local H II
regions, in close agreement with the galaxy center's heliocentric velocity of
$-234$ km s$^{-1}$ \citep{Lu1993}.

\begin{figure*}[htp]
\centerline{\includegraphics[angle=0,width=17.cm]{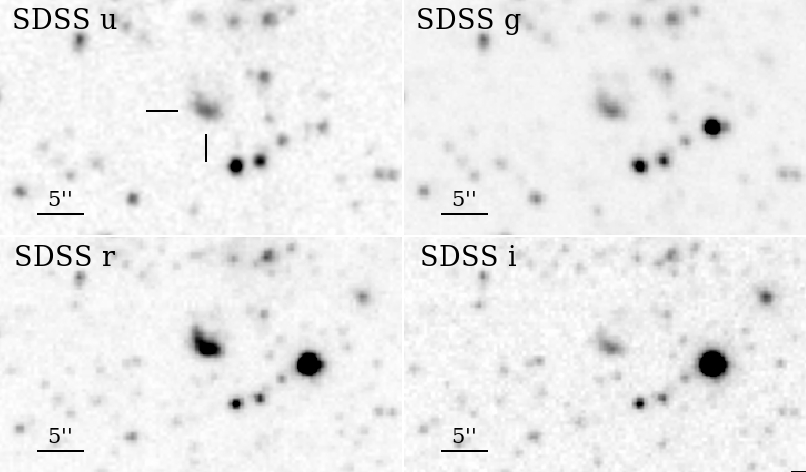}}
\caption{Sloan images of the IC 1613 supernova remnant S8, marked in the u band image. North is up, East is to the left.}
\label{Sloan}
\end{figure*}

In addition, we obtained moderate dispersion spectra (R $\simeq$12,400) of the
S8 remnant in January 2020 with the MDM 1.3m McGraw-Hill telescope using a
Boller \& Chivens Spectrograph (CCDS) in order to investigate the remnant's
velocity structure in finer detail.  This spectrograph uses a Loral 1200
$\times$ 800 CCD detector. A 1800 grooves mm$^{-1}$ grating blazed at 4700 \AA
\ was used to yield a wavelength coverage of 330 \AA.  Three $1500$ seconds
exposures were taken centered on the remnant's H$\alpha$ emission  under 1.5 --
2.0 arcsec seeing conditions.  With a spectral scale of 0.265 \AA \ per pixel
and a measured comparison lamp resolution of 2.0 pixels using a 1.5 arcsec wide
slit, this set-up gave a velocity resolution of $\simeq$24 km s$^{-1}$ at
H$\alpha$.  This resolution was viewed adequate to explore the remnant's
H$\alpha$ emission velocity considering the remnant's reported 250 to 300 km
s$^{-1}$ expansion \citep{Lozinskaya1998,Lozinskaya2009}.

Standard pipeline data reduction of both images and spectra made use of AstroPy
and PYRAF\footnote{PYRAF is a product of the Space Telescope Science Institute,
which is operated by AURA for NASA.}. None of the October 2019 or January 2020
nights were strictly photometric, with occasional light and variable cirrus.
Spectra were reduced using the software L.A.\ Cosmic \citep{vanDokkum2001} to
remove cosmic rays.  Spectra were calibrated using Ne, Xe, and Ar lamps and
spectroscopic standard stars \citep{Oke1974,Massey1990}.

\section{Results}
\subsection{Image Data}

As shown in Figure~\ref{WideFOV}, the S8 remnant lies at the southern edge of a
cluster of H II regions and large emission shells located in the northeastern
outskirts of the IC 1613 galaxy. S8 is the only know SNR in the galaxy and has
the highest surface brightness of any emission line nebula in IC 1613
\citep{Lozinskaya1998}. IC~1613 is located at a fairly high Galactic latitude
($-60.6\degr$) with a low estimated foreground Galactic extinction ($E(B-V)
= 0.025$; \citealt{Schlafly2011}). 

Although several {\sl HST} observing programs imaged parts of IC~1613, none
covered the S8 site which might have informed us about the remnant's
fine-scale emission structure.  Outside of Sandage's initial H$\alpha$ image,
there are also few high-resolution images of the S8 the remnant in the literature (e.g.,
\citealt{Lozinskaya1998,Rosado2001}).  Consequently, we investigated the remnant's
appearance both through archival optical survey data plus our new imaging data.

Broadband images of the S8 remnant are shown in Figure~\ref{Sloan} where we present
Sloan u,g,r,i images. In these images, the remnant exhibits a distinct crescent shape
$3\farcs5 \times 5\farcs0$ in size toward the southeast, consistent with the size cited by \citet{Sandage1971}. 
The crescent is significantly brighter in its
southwest corner, except in the u band where it appears more uniform in brightness.
Our V band and F675W images matches that seen in the
Sloan g and r images.

\begin{figure*}[htp]
\centerline{\includegraphics[angle=0,width=18.cm]{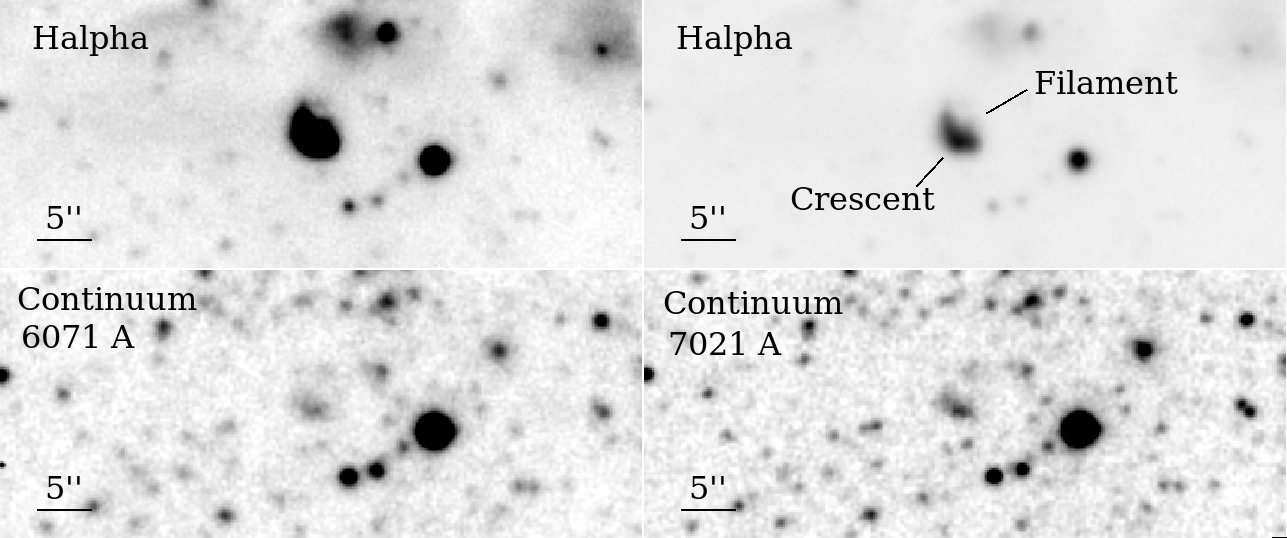}}
\caption{
Top Panels: H$\alpha$ images of the S8 remnant; left panel shows its full
extent in H$\alpha$ while the right panel presents a lower contrast version 
which better reveals the overall morphology including a single, 
interior filament.
Lower Panels: Red continuum images ($\lambda_{\rm c}$ = 6071 and 7021 \AA) 
showing the presence of continuum emission with a morphology similar to that of the remnant's H$\alpha$ emission. North is up, East is to the left.
}
\label{Ha_images}
\end{figure*}

With the exception of the Sloan r band image, the remnant's total flux is
remarkably steady in the u,g, and i band images suggesting a relatively flat
SED.  Estimated remnant ugri magnitudes of 18.6, 18.7, 17.5, and 18.8 mag ($\pm
0.15$ mag) using neighboring stars in the Sloan DR16 database support this
conclusion.  S8's bright appearance in the r band is likely due to that
filter's sensitivity to the remnant's strong H$\alpha$, [\ion{O}{1}] 6300, 6364
\AA, [\ion{N}{2}] 6548,6583 \AA, and [\ion{S}{2}] 6716, 6731 \AA \ line emissions. 

However, the remnant's bright appearance in Sloan i band is surprising.
Previous spectra of the SNR along with our own spectra (see \S3.2 below) show that S8
emits no strong emission lines in the Sloan i filter's $\sim7000 - 8500$ \AA \
bandpass \citep{Peimbert1988,Lozinskaya1998}. The same is also true
for the Sloan u band image where the filter's bandpass is sensitive only to the
remnant's only modest strength [\ion{O}{2}] 3726, 3729 \AA \ emission.

Typically, evolved SNRs are mainly emission line nebulae and do not exhibit
significant continuum emission, with the Crab Nebula being one of the few and most
notable exceptions.  Thus, the detection of the S8 remnant in the Sloan u and i
band images was unexpected. This led us to obtain several narrow passband filter images
to explore the remnant's continuum emission and morphology.  

In Figure~\ref{Ha_images}, we present our sub-arcsecond H$\alpha$ and continuum images of S8,
with the upper panels showing the remnant's H$\alpha$ emission while the lower
panels show the remnant's continuum emission structure.  The top left-hand panel
shows the remnant's full H$\alpha$ emission extent where it appears
as a bright and sharply defined, thick crescent shaped nebula, along with
possible faint diffuse extension to the northeast. Its elliptical shape has
angular dimensions $\simeq 3\farcs5 \times 5\farcs0$, with a major axis at PA =
$\simeq30^{\rm o}$.  These numbers are considerably smaller than the $6\arcsec
\times 8\arcsec$ cited by \citet{Lozinskaya1998} and \citet{Rosado2001} but in good
agreement with the major diameter of $5\farcs4$ reported by \citet{Sandage1971} and
\citet{Peimbert1988}. An angular size of a $3\farcs5
\times 5\farcs0$  translates to 12 pc x 18 pc at 725 kpc. 

The top right panel shows a lower contrast version of this same H$\alpha$ image
emphasizing the elongated crescent shape of its brightest emission, along with
a small, $\sim1\farcs5$ long filament-like feature sticking out of its northern
boundary. An [\ion{O}{3}] filter image (not shown) presents a nearly identical
size and morphology to that seen in H$\alpha$, consistent with previous emission
line image results \citep{Lozinskaya1998}.

Both red continuum filter images, centered at 6071 and 7021 \AA, reveal
coincident and unexpectedly bright continuum nebulosity exhibiting a size and
morphology like that seen for the remnant's line emissions.  This continuum
emission is best resolved in the 7021 \AA \ filter image (FWHM = $0\farcs82$).
This continuum emission appears diffuse with no clear evidence for the presence
of one or more point sources.  Consistent with the Sloan images, the remnant's
continuum is brightest toward the southwest and we estimate the peak brightness
in the 6071 \AA\ continuum filter to be $\sim$ 4-5\% that of H$\alpha$.

\subsection{Low-Dispersion Spectra}
\subsubsection{Optical Spectral Properties of S8}

\begin{figure*}[t]
\centering
\includegraphics[width=0.95\textwidth]{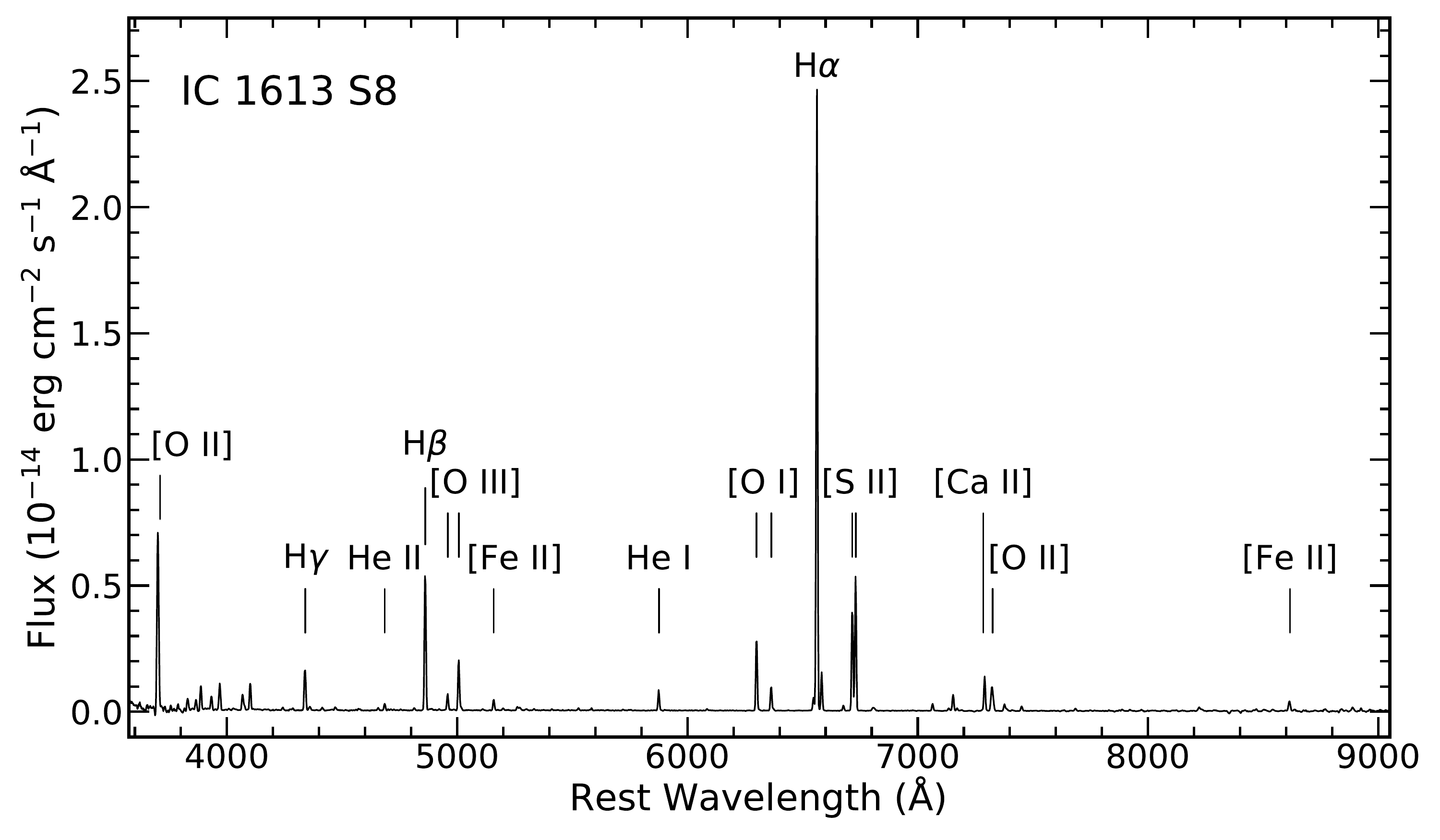}
\caption{Observed optical spectrum of the S8 remnant.}
\label{full_spectrum}
\end{figure*}

\begin{figure*}[htp]
\centering
\includegraphics[width=0.95\textwidth]{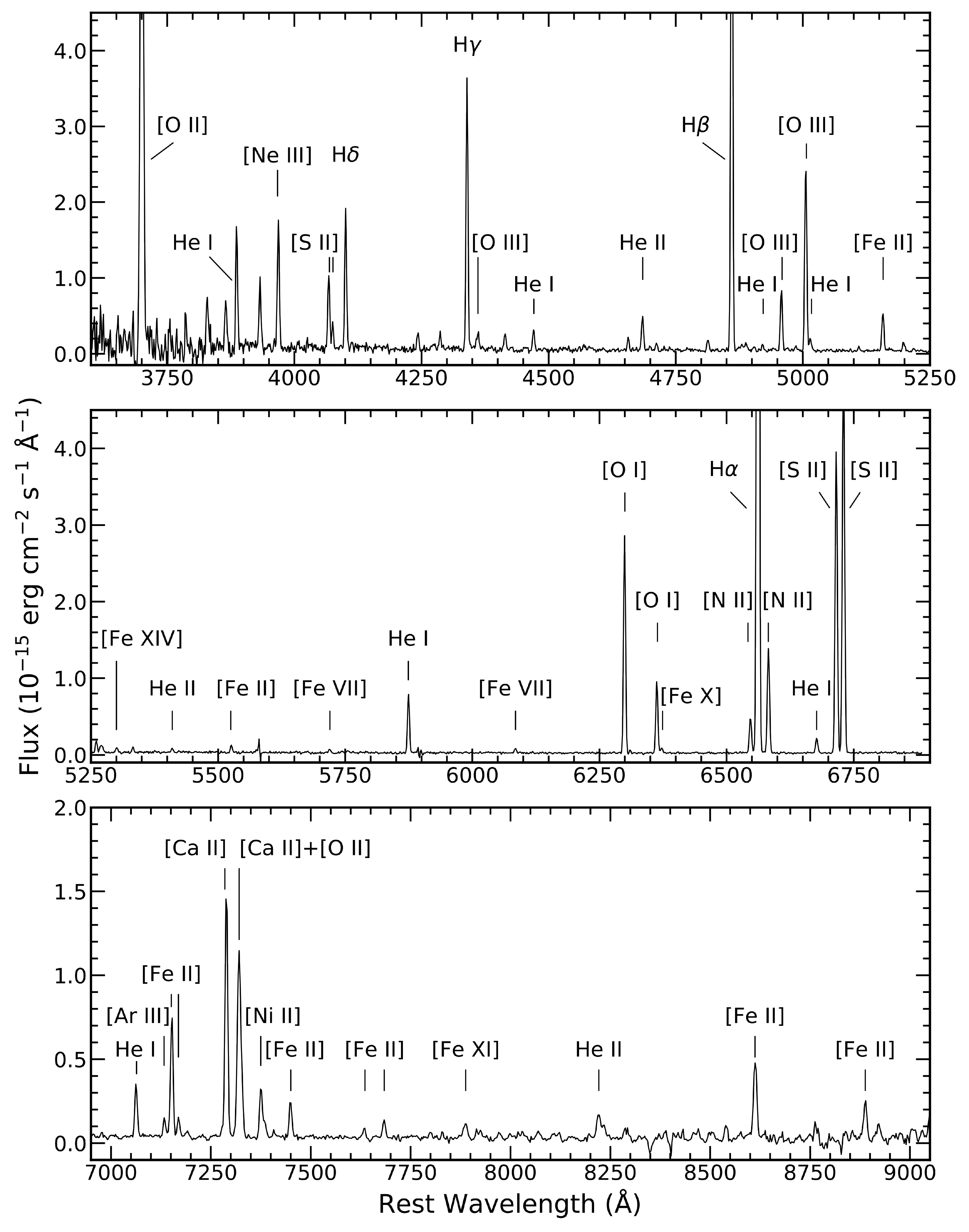}
\caption{Enlargements of the observed spectra of the S8 remnant covering the
3600--5250 \AA, the 5250--6900 \AA, and the 6900--9000 \AA \ regions.
Note the presence of faint [\ion{Fe}{7}] 5721, 6087, [\ion{Fe}{10}] 6374, [\ion{Fe}{11}] 7892,
and [\ion{Fe}{14}] 5303 line emissions. }
\label{3spectra}
\end{figure*}

Our low-dispersion spectrum of the S8 SNR covering the wavelength range of 3600
to 9000 \AA \ is shown in Figure~\ref{full_spectrum}, with enlarged plots shown in Figure~\ref{3spectra}.
These figures show the remnant's full optical spectrum clearly for the first time,
with as good or better S/N and spectral resolution than 
previously available. (Note: Faint continuum emission was detected but is too weak to see in this plot.)

Our spectrum is in general agreement with earlier published data and this is
reflected in Table~\ref{Table_1}, where we compare our observed and extinction
corrected relative line intensities for most of the stronger lines with those
of previous works.  Columns 1 and 2 list the observed line strengths for the S8
remnant from \citet{DD1983} and \citet{Lozinskaya1998}, our observed and
extinction corrected values in Columns 3 and 4, and extinction corrected line
strengths from \citet{Peimbert1988} and \citet{Dopita2019} in Columns 5 \& 6.
We chose to list relative line strengths from previous and our own 
measurements in this way, to more easily show the consistency of our results with prior works for both the SNR's main emission lines
(Cols.\ 1 - 3) and the major and minor
emission lines in the longer line lists of \citet{Peimbert1988} and
\citet{Dopita2019} (Cols.\ 4 - 6).

IC 1613 has a low amount of foreground and internal reddening
($E(B-V)$ = 0.02; \citealt{Saha1992,Lee1993}), with H~II region spectral
observations indicating $E(B-V)$ = $0.10$ \citep{Lee2003}.  Based on our
observed H$\alpha$/H$\beta$ ratio of 4.20, we calculated an $E(B-V)$ = 0.34,
equivalent to a value of c $\approx$0.44, assuming an intrinsic ratio of 3.0
and an R value of 3.1 which has been shown to be valid for IC~1613
\citep{Pie2006}. We chose an H$\alpha$/H$\beta$ value greater than the
theoretical value of 2.87 for 10$^{4}$ K due to the likelihood of significant
collisional excitation of the n = 3 level at postshock temperatures seen in
SNRs. Our extinction value is different from that of previous observers, which
themselves differ from one another, and this may reflect real extinction
differences internal to the remnant as suggested by variations in observed
H$\alpha$/H$\beta$ ratios reported by \citet{Lozinskaya1998}.

Even a casual inspection of the S8's emission line strengths makes immediately
obvious the weakness of the metal lines of [\ion{O}{2}] and  [\ion{O}{3}] 4959,
5007 line emissions relative to H$\beta$, the weakness of the [\ion{N}{2}]
6548, 6583 lines, and [\ion{S}{2}] 6716, 6731 relative to H$\alpha$ compared to
Galactic SNRs . In Galactic SNRs as well as those in M31, M33, and other
massive galaxies, [\ion{O}{3}] is usually stronger and often much stronger than
H$\beta$, with the [\ion{N}{2}] 6583 line usually comparable to that of
H$\alpha$. However in S8 these lines are relatively weak.  The S8 spectrum
carries clear signs of low metal ISM abundances like that seen in the
Magellanic Clouds, not surprising given IC~1613's estimated metallicity of
$0.04 - 0.13$ Z$_{\odot}$
\citep{Kingsburgh1995,Peimbert1988,Skillman2003,Garcia2014,Berger2018}. In
Table~\ref{Table_1}, we also include lists of published relative emission line
strengths for two young core-collapse LMC remnants, namely N49 and N63A, which
are discussed in relation to S8 in $\S$4.2.

\begin{deluxetable*}{lcccccccc}[ht]
\tiny
\tablecolumns{9}
\tablecaption{Emission Line Strengths of S8 Compared to Previous Results and the LMC SNRs N49 and N63A}
\tablehead{\colhead{} & \multicolumn{6}{c}{\underline{~~~~~~~~~~~~~~~~~~~~~~~~~~~~~~~~~~~~~~~S8 ~~~~~~~~~~~~~~~~~~~~~~~~~~~~~~~~~~~~~~~~~~}}  & \colhead{\underline{N49}} & \colhead{\underline{N63A}}  \\
   \colhead{Emission Line}  &  \colhead{DD83} & \colhead{L98} & \colhead{this work}  & \colhead{this work} & \colhead{Peimbert88} & \colhead{Dopita19} & \colhead{Dopita19} & \colhead{RD90} \\
    \colhead{(\AA)} & \colhead{F($\lambda$)} & \colhead{F($\lambda$)} & \colhead{F($\lambda$)} &  \colhead{I($\lambda$)} & \colhead{I($\lambda$)} & \colhead{I($\lambda$)} &
     \colhead{I($\lambda$)} & \colhead{I($\lambda$)} }
\startdata
 $[$\ion{O}{2}$]$ 3727               &  107.0  & \nodata &  125    &  168    &  136.0  & 148.2  & 562.4 &  368.9    \\
 $[$\ion{Ne}{3}$]$ 3869              & \nodata & \nodata &  5.6    &  7.6    &   8.1   &   9.1  &  37.0 &  11.8     \\
  \ion{He}{1} + \ion{H}{1} 3889      & \nodata & \nodata &  13.8   &  18.7   &  18.6   &  19.0  &  22.1 &  18.6     \\
 $[$\ion{Ca}{2}$]$ 3934              & \nodata & \nodata &  8.0    &  11.5   &  10.0   &  10.6  &  15.6 &   8.2     \\
 $[$\ion{Ne}{3}$]$ 3967              & \nodata & \nodata &  14.0   &  18.0   &  24.0   &  22.6  &  33.3 &   24.5    \\
 $[$\ion{S}{2}$]$ 4068, 4076         &  15.0   & \nodata &  11.6   &  17.1   &  17.8   &   17.7 &  39.3 &   29.1    \\
  H$\delta$   4102                   & 18.0    & \nodata &  16.1   & 21.2    &  25.1   &  24.0  &  25.2 &   26.6    \\
  H$\gamma$   4340                   & 35.0    &  35.0   &  32.8   &  38.9   &  45.7   &  46.1  &  45.4 &   44.4    \\
 $[$\ion{O}{3}$]$ 4363               &   2.0   &   5.0   &   2.3   &  2.7    &  4.0    &   3.1  &   3.9 & \nodata   \\
 \ion{He}{1} 4472                    & \nodata &   2.0   &   3.0   & 3.5     &  4.0    &   3.1  &   4.1 &    2.9    \\
 \ion{He}{2} 4686                    &   3.0   &    4.8  &   4.5   &  5.0    &  5.0    &   5.9  &   5.4 &    2.1    \\
  H$\beta$    4861                   & 100.0   &   100.0 & 100.0   & 100.0   &  100.0  &  100.0 & 100.0 &  100.0    \\
 \ion{He}{1} 4922                    & \nodata & \nodata &   0.5   & 0.5     & \nodata &   1.0  &   1.2 & \nodata   \\
 $[$\ion{O}{3}$]$ 4959               &   12.0  & \nodata &  11.3   & 11.2    & 12.0    &   11.8 &  25.5 &   27.8    \\
 $[$\ion{O}{3}$]$ 5007               &   35.0  &   37.3  &  37.8   & 35.9    & 37.0    &   35.3 &  78.3 &   81.0    \\
 \ion{He}{1} 5015                    & \nodata & \nodata &   2.1   &  1.8    & \nodata &   0.9  &   1.2 & \nodata   \\
 $[$\ion{N}{1}$]$ 5200               &    2.0  &    2.8  &   1.5   &  1.4    & \nodata &   1.4  &   0.7 &    6.1    \\
 \ion{He}{1} 5876                    &   12.0  &    11.9 &  14.0   & 11.0    &   10.0  &  10.9  &  11.2 &   10.9    \\
 $[$\ion{O}{1}$]$ 6300               &   45.0  &    68.0 &  49.7   & 37.7    &   34.0  &   38.2 &  90.0 &  119.8    \\
 $[$\ion{S}{3}$]$ 6312               & \nodata & \nodata &   0.7   &  0.6    & \nodata &   1.0  &   1.0 & \nodata   \\
 $[$\ion{O}{1}$]$ 6364               & \nodata & \nodata &  18.4   & 12.9    &   9.0   &  13.3  &  30.9 &   38.8    \\
 $[$\ion{N}{2}$]$ 6548               & \nodata & \nodata &  10.3   & 7.0     & \nodata &   7.1  &  21.9 & \nodata   \\
  H$\alpha$   6563                   &  417.0  &   395.0 &  420.0  & 300.0   & 295.0   &  331.0 & 301.1 &  437.9    \\
 $[$\ion{N}{2}$]$ 6583               &   32.0  &    31.1 &   27.3  & 18.3    &  21.0   &   20.3 &  59.2 &  112.8    \\
 \ion{He}{1} 6678                    & \nodata & \nodata &   3.4   &  2.4    & \nodata &   2.8  &   2.8 &    4.9    \\
 $[$\ion{S}{2}$]$ 6716               &   60.0  &    64.5 &   73.0  & 50.9    & 51.0    &   54.5 &  85.1 &  109.3    \\
 $[$\ion{S}{2}$]$ 6731               &   74.0  &    79.6 &   96.0  & 67.0    & 65.0    &   71.1 & 120.6 &  142.7    \\
 \ion{He}{1} 7065                    & \nodata & \nodata &   5.0   &  3.3    & \nodata &   3.7  &   4.3 &    3.8    \\
 $[$\ion{Ar}{3}$]$ 7136              & \nodata & \nodata &    2.1  &  1.2    & \nodata &   1.4  &   6.3 &    6.8    \\
 $[$\ion{Ca}{2}$]$ 7291              &   24.0  &   14.7  &   24.7  & 15.9    &  20.0   &   19.0 &  26.9 &   14.0    \\
 $[$\ion{O}{2}$]$ 7319 + $[$\ion{Ca}{2}$]$ 7325 &16.0 &13.3& 25.9  & 16.6    &  22.0   &   17.5 &  54.1 &   34.1    \\
 $[$\ion{Ni}{2}$]$ 7378              & \nodata & \nodata &    5.6  &  3.6    & \nodata &    3.4 &   9.3 &    5.4    \\
 $[$\ion{Ni}{2}$]$ 7411              & \nodata & \nodata &    0.6  &  0.4    & \nodata &    0.6 &\nodata& \nodata   \\
 $[$\ion{Cr}{2}$]$ 8230              & \nodata & \nodata &    3.0  &  1.8    & \nodata &   1.7  &\nodata& \nodata   \\
  \ion{He}{2}  8237                  & \nodata & \nodata &    1.5  &  0.7    & \nodata & \nodata& 2.5   & 1.0       \\
                                     &         &         &         &         &         &        &       &           \\
 $[$\ion{S}{2}$]$ 6716/6731          &  0.81   &  0.80  &   0.76  & 0.76    &  0.78   &  0.77  &  0.71 &   0.76    \\
 $[$\ion{O}{3}$]$ (4959+5007)/4363 &15.4       &$\sim$10  &\nodata& 17.4  &  12.3   &  14.8  & 26.6  & \nodata   \\
 Log I(H$\beta$) erg cm$^{-2}$ s$^{-1}$ &      &  -13.19 &  -13.26 &         & -13.40  &     &       &           \\
\enddata
\tablecomments{Log I(H$\beta$) values listed are uncorrected for reddening.}
\tablerefs{References -- DD83: \citet{DD1983}; L98: \citet{Lozinskaya1998}; Peimbert88: \citet{Peimbert1988}; Dopita19: \citet{Dopita2019}; RD90: \citet{Russell1990} }
\label{Table_1}
\end{deluxetable*}

\begin{deluxetable*}{lccccccc}[htp]
\tiny
\centerwidetable
\tablecolumns{8}
\tablecaption{Relative [\ion{Fe}{2}] Emission Line Strengths in S8, N49, and N63A}
\tablewidth{0pt}
\tablehead{\colhead{}  & \colhead{} & \multicolumn{3}{c}{\underline{~~~~~~~~~~~~~~~~~~~~~S8~~~~~~~~~~~~~~~~~~}} & \multicolumn{2}{c}{\underline{~~~~~~~~~~~ N49 ~~~~~~~~~}} & \colhead{\underline{N63A}} \\
         \colhead{Emission Line}  &  \colhead{Multiplet}     & \colhead{this work}       &  \colhead{this work}   & \colhead{Dopita19}     & \colhead{RS90}       &  \colhead{Vancura92} &  \colhead{RS90} \\
        \colhead{(\AA) }  &  \colhead{Number}    & \colhead{ F($\lambda$)}  & \colhead{I($\lambda$)}  & \colhead{I($\lambda$)} &  \colhead{I($\lambda$)}  & \colhead{I($\lambda$)} & \colhead{I($\lambda$)}  }
\startdata
 $[$\ion{Fe}{2}$]$ 4244, 4245 & 21F     &   2.0   & 2.3  &  2.6  &  4.0   & \nodata & 1.6      \\
 $[$\ion{Fe}{2}$]$ 4287       & 7F      &   1.2   & 1.5  &  2.1  &  3.6   & \nodata & \nodata  \\
 $[$\ion{Fe}{2}$]$ 4358, 4359 & 21F, 7F &   1.4   & 1.7  &\nodata& 13.5   & \nodata & \nodata  \\
 $[$\ion{Fe}{2}$]$ 4414       & 7F      &   2.5   & 2.9  &  3.1  &  3.9   & \nodata & \nodata  \\
 $[$\ion{Fe}{3}$]$ 4658       & 3F      &   2.0   & 2.3  &  1.7  &  3.9   & \nodata & \nodata  \\
 $[$\ion{Fe}{3}$]$ 4814       & 3F      &   1.2   & 1.4  &  1.7  &  2.3   & \nodata & 2.8      \\
 $[$\ion{Fe}{3}$]$ 4882       & 2F      &   0.5   & 0.5  &  1.0  &\nodata & \nodata & \nodata  \\
 $[$\ion{Fe}{2}$]$ 4890       & 4F, 3F  &   0.6   & 0.6  &\nodata&\nodata & \nodata & \nodata  \\
 $[$\ion{Fe}{2}$]$ 5111       &  19F    &   0.5   & 0.4  &  0.9  &  0.8   & \nodata & 1.6      \\
 $[$\ion{Fe}{2}$]$ + $[$\ion{Fe}{7}$]$ 5159 & 18F, 19F, 2F&7.7&6.9&6.8& 10.2&  10.0 & 5.1      \\
 $[$\ion{Fe}{2}$]$ 5220       & 19F     &   0.7   &  0.6 &  0.6  & \nodata & \nodata& \nodata  \\
 $[$\ion{Fe}{2}$]$ 5262       & 19F     &   3.4   &  2.7 &  2.5  &  4.8    & 3.5    & 2.1      \\
 $[$\ion{Fe}{2}$]$ 5273       & 18F     &   2.6   &  2.3 &  0.4  & \nodata & 1.3    & \nodata  \\
 $[$\ion{Fe}{2}$]$ 5334       & 19F     &   1.2   &  0.8 &  0.9  & \nodata &  1.9   & \nodata  \\
 $[$\ion{Fe}{2}$]$ 5527       & 17F     &   1.6   &  1.3 &  1.1  &  2.0    &  2.1   & \nodata  \\
 $[$\ion{Fe}{2}$]$ 7155       & 14F     &   12.3  &  7.8 &  8.5  &  16.9   & 12.6   & 7.5      \\
 $[$\ion{Fe}{2}$]$ 7172       & 14F     &    1.9  &  1.1 &  1.0  &   4.1   &  2.1   & \nodata  \\
 $[$\ion{Fe}{2}$]$ 7387       & 14F     &    2.5  &  1.8 &  1.2  & \nodata & \nodata& \nodata \\
 $[$\ion{Fe}{2}$]$ 7453       & 14F     &    3.6  &  2.3 &  2.8  &   13.9  &  4.1   &   9.6   \\
 $[$\ion{Fe}{2}$]$ 7638       &  1F     &    1.1  &  0.6 &  1.1  &   6.6   &  1.5   &   2.6   \\
 $[$\ion{Fe}{2}$]$ 7686       & 14F     &    1.9  &  1.0 &  0.8  &   2.1   &  1.5   &   1.8    \\
 $[$\ion{Fe}{2}$]$ 8617       & 13F     &    9.2  &  4.3 &  9.7  &  19.9   & 13.5   & 12.1    \\
 $[$\ion{Fe}{2}$]$ 8892       & 13F     &    4.4  &  1.9 &  3.1  &   5.9   &  4.1   & 6.0     \\
\enddata
\tablecomments{Listed line strengths are relative to H$\beta$ = 100.}
\tablerefs{Dopita19: \citet{Dopita2019}; RS90: \citet{Russell1990}; Vancura92: \citet{Vancura1992}  }
\label{Table_2}
\end{deluxetable*}

\subsubsection{Electron Densities and Temperatures}

The ratio of the [\ion{S}{2}] 6716, 6731 lines can be used to estimate the electron
density in the S$^{+}$ recombination zone and is nearly independent of electron
temperature \citep{Osterbrock2006}. As shown in Table~\ref{Table_1}, both
\citet{DD1983} and \citet{Lozinskaya1998} reported the 6716/6713 ratio
$\approx$ 0.80, whereas \citet{Peimbert1988} and \citet{Dopita2019} find
somewhat lower values of 0.78 and 0.77, respectively.  While
\citet{Peimbert1988} estimated an electron density, $n_{e}$, of $1500 \pm 230$
cm$^{-3}$.  \citet{Lozinskaya1998} estimated of $\simeq$ 1300 cm$^{-3}$ based
on an average ratio of $0.80 \pm0.05$ and using atomic level calculations by
\citet{DeRobertis1987}.

From our higher resolution blue grism spectra, we find a 6716/6731 ratio of
$0.76 \pm0.013$ which suggests an $n_{e}$, of $1600$ cm$^{-3}$. However,
if we adopt the revised formulation of \citet{Proxauf2014} which used CLOUDY
models \citep{Ferland2013} to calibrate the density sensitive [\ion{S}{2}]
6716/6731 line ratio, this observed ratio translates into a electron density,
$n_{e}$, of $1060 \pm100$ cm$^{-3}$. As noted by \citet{Proxauf2014}, their
model predicts lower densities by some 20\% from those of previous works.
However, due to the low metallicity of the IC 1613 galaxy, this value must be
revised up a bit to $\simeq$ 1200 cm$^{-3}$ \citep{Kewley2019}.

Considering possible differing slit placements by us and previous observers,
the agreement for the [\ion{S}{2}] 6716/6731 ratio between observers suggests
little in the way of density variations across the S8 remnant. However,
based on observed differences of the [\ion{S}{2}] 6716/6731 ratio across
the remnant, \citet{Lozinskaya1998} claim density variations of 400 - 600
cm$^{-3}$, along with a single dense knot with n$_{e}$ $\sim$ 2400 cm$^{-3}$
along the remnant's northern edge.

\begin{deluxetable*}{lcccccc}[ht]
\tiny
\centerwidetable
\tablecolumns{7}
\tablecaption{Optical Coronal Line Emissions in S8 and Young LMC SNRs  }
\tablewidth{0pt}
\tablehead{\colhead{Emission}   &  \multicolumn{3}{c}{\underline{~~~~~~~~~~~~~~~~~~S8 ~~~~~~~~~~~~~~~~~~~~~}}        &  \colhead{\underline{N49}}  &  \colhead{\underline{N63A}}  \\
           \colhead{Line}  &  \colhead{this work}      & \colhead{this work}   & \colhead{Dopita19}      & \colhead{Dopita19}    & \colhead{RD90}         \\
        \colhead{(\AA \ )} &  \colhead{F($\lambda$)} & \colhead{I($\lambda$)} & \colhead{I($\lambda$)}  & \colhead{I($\lambda$)} & \colhead{I($\lambda$)}  }
\startdata
 $[$\ion{Fe}{7}$]$ 5720    &   0.7   &  0.5    &   0.3  & \nodata & \nodata  \\
 $[$\ion{Fe}{7}$]$ 6087    &   0.9   &  0.8    &   1.4  &  0.4    & \nodata  \\
 $[$\ion{Fe}{10}$]$ 6374   &   1.2   &  0.9    &   0.5  &  1.1    & \nodata  \\
 $[$\ion{Fe}{11}$]$ 7892   &   2.1   &  1.0    &   1.0  &  1.8\tablenotemark{a}    &    0.3   \\
 $[$\ion{Fe}{14}$]$ 5303   &   1.1   &  0.8    &   1.9  &  1.2    & \nodata  \\
\enddata
\tablecomments{Listed line strengths are relative to H$\beta$ = 100.}
\tablenotetext{a}{N49 line value above 7000 \AA \ is taken from \citet{Russell1990}.}
\tablerefs{Dopita19: \citet{Dopita2019}; RD90: \citet{Russell1990}}
\label{Table_3}
\end{deluxetable*}

The remnant's electron temperature of its O$^{+2}$ emitting regions can be
calculated from the ratio, $R$, equal to I(4959 + 5007)/I(4363)
\citep{Osterbrock2006}.  If the electron density is $<< 10^{6}$ cm$^{-3}$, then
$R = 7.90 \times$ e$^{3.29 \times 10^{4}/T_{e}}$. For electron temperatures
below $50 \times 10^{3}$ K, the [\ion{O}{3}] 4363 \AA\ line is relatively weak
relative to the 5007 \AA \ line.  

In the case of S8, where the [\ion{O}{3}] 4959, 5007 \AA\ lines are not especially
strong,  estimating an electron temperature for the O$^{+2}$ based on the much
weaker 4363 \AA \ line requires especially good S/N spectra.  In addition, as noted
by several authors 
measuring the [\ion{O}{3}] electron temperature can be complicated by
blending from [\ion{Fe}{2}] emissions around 4360 \AA \  with the temperature sensitive
[\ion{O}{3}] 4363 \AA \ line \citep{Osterbrock1973,Fesen1982a,Peimbert1988,Curti2017}. Given the numerous [\ion{Fe}{2}] lines in the S8
spectrum, blending of [\ion{O}{3}] 4363 \AA \ emission with that of [\ion{Fe}{2}]
emission lines thus poses a concern for obtaining an accurate [\ion{O}{3}]
electron temperature for the S8 remnant.

Blending of [\ion{Fe}{2}] emission with  [\ion{O}{3}] 4363 \AA \ emission may
have lead to artificially small $R$ values in past studies and hence relatively
high estimated [\ion{O}{3}] temperatures for S8.  \citet{DD1983} cite a value
of 15.4 indicating T$_{e}$ $\simeq 50 \times 10^{3}$, while
\citet{Peimbert1988} find $R$ = 12.3 leading to their estimate of T$_{e}$ =
$(80 \pm 15) \times 10^{3}$ K.  While updated atomic constants would lower this
T$_{e}$ estimate closer to $75 \times 10^{3}$ K, this would still be an
exceptionally high temperature for a completely radiative SNR filament (see
\citealt{Fesen1982a,FH1996}).  \citet{Lozinskaya1998} do not list a value
for $R$ but their data suggests an even lower value $\sim$10 implying T$_{e} >$
10$^{5}$ K.  An  $R$ value of 14.8  cited by \citet{Dopita2019} implies T$_{e}$
$\simeq$53,000 K which is still relatively hot for a radiative shock filament.

Our blue grism spectrum of S8 is of sufficient spectral resolution (3.5 \AA) to
show a weak blended emission feature around 4361 \AA.  There are potentially
three [\ion{Fe}{2}] emission lines which could contribute to a blended feature.
Their lab wavelengths and forbidden multiplet numbers are: 4358.10 (6F),
4358.37 (21F), and 4359.34 (7F). Analysis of the observed blended feature
suggests just two components near 4359.1 and 4362.6 \AA \ which we attribute to
the [\ion{Fe}{2}] 4359.34 line and the [\ion{O}{3}] 4363.21 line, respectively.
This conclusion is supported by calculated [\ion{Fe}{2}] transition
probabilities \citep{Garstang1962}, which predicts the 4359.34 (7F) to be
stronger than the 4358.37 (21F) line.  From deblended extinction corrected
intensities relative to H$\beta$, we find the ratio of the [\ion{O}{3}] lines
(4959 + 5007)/4363 $= 17.4 \pm1.4$ indicating T$_{e} = 41,500 \pm4000$ K for
the O$^{+2}$ zone.  

\subsubsection{Forbidden Iron Emission Lines}

The S8 remnant displays a wealth of forbidden Fe lines, both low ionization
[\ion{Fe}{2}] lines and much higher ionization coronal lines such as [\ion{Fe}{10}]
and [\ion{Fe}{14}] (see Fig.~\ref{3spectra}).  Our observed and extinction corrected line
strengths of [\ion{Fe}{2}] and [\ion{Fe}{3}] lines are listed in Table~\ref{Table_2}, along
with extinction corrected values taken from \citet{Dopita2019}. As seen in
this table, with the exception of lines above 8000 \AA, our measurements and
those of \citet{Dopita2019} for these forbidden Fe lines are in very good agreement.

Interestingly, the spectrum of S8 also shows the presence of several coronal emission lines of iron
(Fig.~\ref{full_spectrum}).  These include include [\ion{Fe}{10}] 6374, [\ion{Fe}{11}] 7892, and
[\ion{Fe}{14}] 5303.  Although the [\ion{Fe}{7}] 5720, 6087 are
not always included in discussions of iron coronal lines, we include these here
since they represent a higher degree of ionization ($>$99 eV) than the lines
listed in Tables~\ref{Table_1} and \ref{Table_2}.

The detection of optical coronal lines of iron in the spectrum of the S8
was only recently reported \citep{Dopita2019}. Although we were initially
unaware of these results when we obtained our spectra, we confirm their
basic results except for the issue of [\ion{Fe}{9}] as we discuss below.  Coronal
line emissions from S8 are somewhat unexpected in view of both the weakness of [\ion{O}{3}]
emission and the extensive number of [\ion{Fe}{2}] lines seen in the spectrum
(Table~\ref{Table_2}) and indicates a wider range of shock velocities present
in the remnant than previously realized.  Relative line strengths, both observed and extinction
corrected coronal line emissions are shown in Table~\ref{Table_3} along with those reported by 
\citet{Dopita2019}.

\begin{figure*}[ht]
\centering
\includegraphics[width=0.95\textwidth]{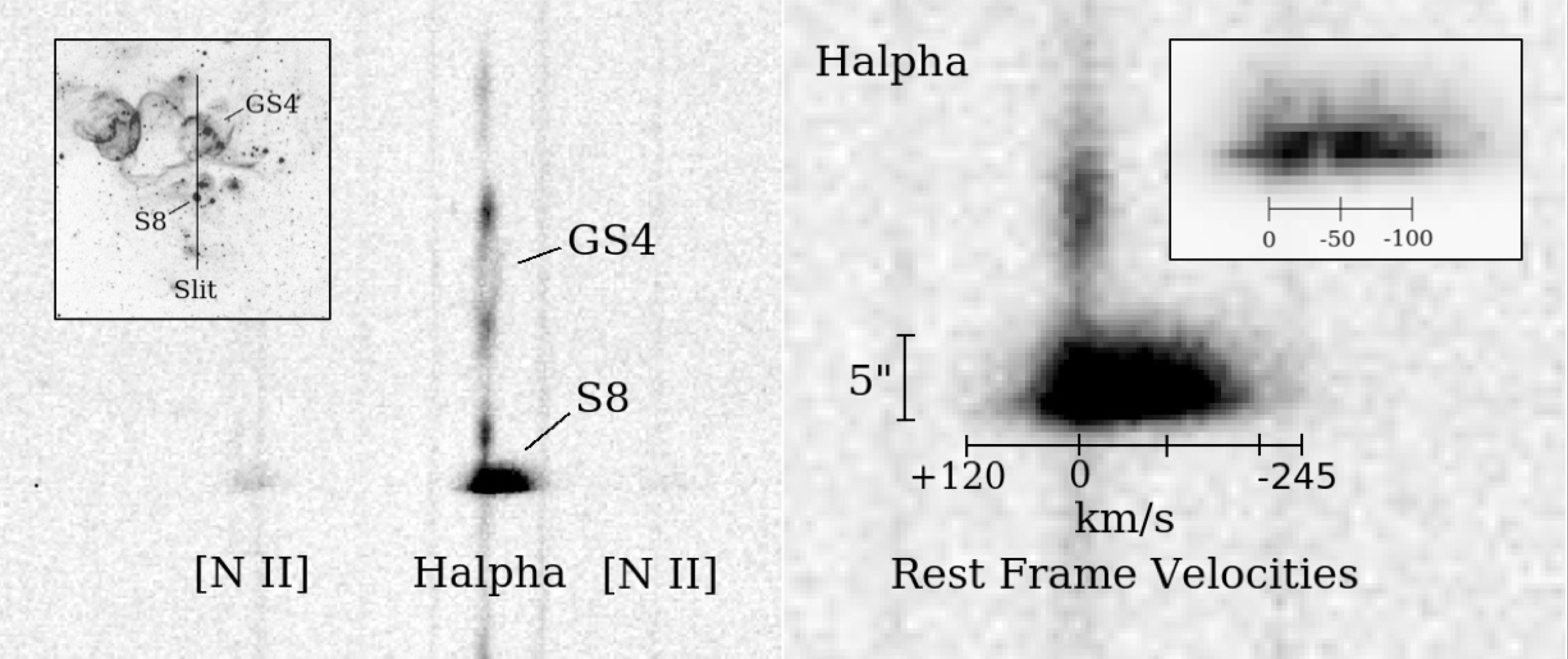}
\caption{Left panel: 2D moderate resolution spectrum (red to the left, blue to the right) showing H$\alpha$ emissions
from S8 and neighboring H II regions. Insert image shows location of the North-South slit
crossing over the S8 remnant and the superbubble GS4.  North is up, East to the left.
The narrow vertical H$\alpha$ emission from the neighboring H II regions was used
to establish the rest frame velocity for S8.  Right panel: A blow-up of S8's H$\alpha$ emission
showing the remnant's internal radial velocity structure, its velocity range, and a central cavity.
Small insert image shows the 45 km s$^{-1}$ wide emission cavity. }
\label{velocity}
\end{figure*}

The presence of [\ion{Fe}{10}], [\ion{Fe}{11}], and [\ion{Fe}{14}] lines, albeit weak,
requires postshock plasma temperatures between 1.2--$2.0 \times 10^{6}$
K \citep{Nussbaumer1970}. Because the maximum temperature just behind a shock with velocity, V,
is approximately
\begin{equation}
 {\rm T_{max}} \simeq \frac{3}{32} \frac{\rm{m_H} \rm{V^{2}}}{\rm k}
\end{equation}
postshock temperatures $\sim2 \times 10^{6}$ K indicate the presence of shocks with velocities 
$\simeq$350 km s$^{-1}$ (see Fig.~\ref{3spectra}). This estimate is supported by the observed line ratio of the [\ion{Fe}{14}] 5303/[\ion{Fe}{11}] 7892 of 0.80 (our data)
and 0.52 from \citet{Dopita2019} (see Table~\ref{Table_3}) which suggests a value of $\sim$330 km s$^{-1}$ based on computed 5303/7892 line ratios as a 
function of blast wave velocity \citep{Vogt2017}.

While coronal line emissions are unusual in SNR spectra in general, they have
been seen in a few young LMC remnants, namely N49, N63A, and N103b
\citep{DM1979,Dopita2019}, and a few especially hot [\ion{O}{3}] filaments in the
Cygnus Loop which has a blast wave velocity around
350 km s$^{-1}$ \citep{Sau1990,FH1996,Raymond2015}.
\citet{Keenan1987} showed that the [\ion{Fe}{7}] 5159/6087 line strength
ratio is a function of temperature and density, but the 5159 \AA \ emission
line is too blended with the [\ion{Fe}{2}] lines at 5158 and 5189 to allow an
estimate of  [\ion{Fe}{7}] gas temperature.

Although we find S8's coronal line strengths generally in agreement with those
reported by \citet{Dopita2019}, we do not support their claimed detection of
[\ion{Fe}{9}] lines at 4359.1 and 8234.5 \AA \ both in S8 and in some LMC SNRs.
If this were true, it would mark the first finding of [\ion{Fe}{9}] line
emissions in any SNR (e.g., \citealt{FH1996}).  Moreover, to our knowledge no
[\ion{Fe}{9}] lines have ever been reported detected in AGNs where coronal
lines are commonly seen \citep{Nussbaumer1970,Ferguson1997,Nazarova1999}, nor
are typically included in general discussions of optical coronal line emissions
from hot plasmas \citep{Graney1990}.

Consequently, we view the presence of either of the two
[\ion{Fe}{9}] lines cited by \citet{Dopita2019} as unlikely in S8's spectrum.  Weak
emission seen at $\simeq$ 4359 \AA \ is more plausibly due to a blend of two
[\ion{Fe}{2}] lines rather than [\ion{Fe}{9}] 4359.1 emission as noted above;
namely 4358.37 (21F) and 4359.34 (7F) \citep{Osterbrock1973}.  Several
[\ion{Fe}{2}] lines from the  7F multiplets are present in the S8 spectrum (see
Table~\ref{Table_2}), making the identification of the feature near 4359 \AA \ as due to
[\ion{Fe}{2}] instead of [\ion{Fe}{9}] likely.

Similarly, their identification of [\ion{Fe}{9}] at 8234.5 \AA \ as the weak
emission feature around 8234 \AA \ also seems doubtful.  In their paper
discussing a spectra of the LMC SNRs N49, N63A, and N103b which exhibit spectra
similar to that of S8, \citet{Russell1990} cite only the detection of
\ion{He}{2} 8236.77 \AA \ around 8230 \AA.  Moreover, in the spectrum of N49 by
\citet{Vancura1992}, they cite the presence of  [\ion{Cr}{2}] 8229.55 \AA \ but
list no feature around  8234 \AA, where  \citet{Levenson1995} find several
[\ion{Cr}{2}] in this region of N63A's spectrum.  Our spectra show a
blended feature with measured centroids at 8221.5 and 8234.2 which we
tentatively attribute to a blend of \ion{He}{2} 8236.8 and  a blend of two
[\ion{Cr}{2}] lines at 8225.2 and 8229.6 \AA.

\subsection{High Dispersion Spectra of S8's H$\alpha$ Emission}

Figure~\ref{velocity} shows a 2D image of our moderate resolution spectrum of S8 and
neighboring H II regions covering the wavelength region around H$\alpha$.  The
small insert image in the left panel shows the location of the long N-S slit on
the sky intersecting both S8 and the large emission shell GS4 located to the
north \citep{Meaburn1988}. Broad H$\alpha$ and [\ion{N}{2}]
emissions can be seen associated with the SNR S8, and much narrower emissions
from H II regions to both the north and south.

The heliocentric radial velocity of IC 1613 is $V_{HEL} = -234$ km s$^{-1}$
\citep{Lu1993}\footnote{Although Sandage is credited with
first identifying S8 as an H~II region, Sandage notes that M.\ Humason obtained a radial
velocity value for IC~1613 of $-238$ km s$^{-1}$ \citep{Humason1955} based on a spectrum
taken of S8 identified as an emission region from a red 200-inch Palomar plate
obtained by W.\ Baade.}.  For the large giant emission shell GS4,
\citet{Meaburn1988},  found $V_{HEL} \simeq -230$ km s$^{-1}$ with an expansion
velocity of $29$ km s$^{-1}$. This expansion velocity is in good agreement with our
measured expansion width of $1.31 \pm0.03$ \AA \ $= 60 \pm 2$  km s$^{-1}$ for
the GS4 shell. In the following discussions, we will adopt GS4's
heliocentric velocity as being the rest velocity for S8's local environment.

The right panel of Figure~\ref{velocity} shows the remnant's full detected radial velocities
cover a range of $\sim365$ km s$^{-1}$, i.e., rest velocities
from $+120$ to  $-245$ km s$^{-1}$, corresponding to $V_{HEL}$ $ = -110$ to $-475$ km
s$^{-1}$.  These maximum and minimum heliocentric velocities are in agreement
with those of \citet{Rosado2001} and of \citet{Lozinskaya1998}, the later who
cite $+145 \pm14$ and $-423 \pm13$ km s$^{-1}$ which they characterize as `high-velocity
features' but which we view as just the velocity extremes of S8's radial
expansion velocities. However, we do not confirm any higher heliocentric
velocities in the range of $-500$ to $-800$ km s$^{-1}$ as claimed by \citet{Lozinskaya1998}.  

As shown in the insert image in the right panel of Figure~\ref{velocity}, the brighter parts of
the remnant H$\alpha$ emission are contained within a fairly small velocity
range; i.e., rest frame velocities, $V_{\rm Rest}$ =  $0$
to $-130$ km s$^{-1}$. The weight of the remnant's emission is clearly blueshifted relative
to the rest frame of the local H II regions.

The brightest H$\alpha$ emission lies in 
two broad brightness maxima centered at rest frame velocities of $-15\pm3$ and
$-65\pm5 $ km s$^{-1}$ and roughly centered on a cavity-like feature $\sim37$ km
s$^{-1}$. The emission is considerably broader on the more blueside of the
H$\alpha$ emission profile extending from $-40$ to $-115$  km s$^{-1}$.
Despite their differences in terms of velocity, both the rear and front 
facing emission maxima exhibit nearly identical peak fluxes.

Both visually and quantitatively, our 2D spectrum presented in Figure~\ref{velocity} agrees
largely with the description of S8's H$\alpha$ emission given by \citet{Rosado2001}, that is one
consisting of two components, one with a FWHM = $66 \pm 10$ km s$^{-1}$ centered
around a rest velocity of $-15$  km s$^{-1}$, with a second and broader
component centered around a rest velocity of $-100$  km s$^{-1}$.  The fact
that the facing, blueshifted emission is noticeably more extended than that of
the rear redshifted emission suggests the possibility of significant internal
extinction between front and rear emission regions, of order A$_{\rm{H}\alpha}$ $\sim
0.5 $ mag.  

An apparent narrow emission depression or `cavity' between these two bright
emission components can also be seen in the right panel's insert of Figure~\ref{velocity}. The emission
decrease is  relatively small in terms of a change in observed H$\alpha$
emission; flux in between the two emission peaks drops only by about 15\%
in intensity, rising up to the two brightness peaks within velocity span of
just $\sim$20 km s$^{-1}$.  The small decrease in flux  of this cavity might be
misleading due to our instrumental 24 km s$^{-1}$ resolution. 
We also note an interesting increase of emission northward along the slit
exactly coincident with the cavity's central velocity.

The reality of this emission cavity is confirmed on the three individual
spectra obtained. It is also supported by the above described findings of
\citet{Rosado2001} and by a similar expansion cavity reported by
\citet{Lozinskaya2009} who find a $75 \pm 25$ km s$^{-1}$ velocity between
front and rear hemispheres. This is larger than the $\simeq 45$  km s$^{-1}$ we
find.  However, this difference may simply reflect that our spectral slit was
not placed in the largest expansion area of the remnant.  Also as noted above,
the more blueshifted maximum emission peak is a broader than the red peak.  If
measured from the velocity average of the broader blueshifted emission rather
than its peak, our data would imply a somewhat larger expansion cavity $\sim55$
km s$^{-1}$.

The remnant's expansion velocity as measured from the
H$\alpha$ emission profile (Fig.~\ref{velocity}) suggests a value around 180 km s$^{-1}$.
However, if measured from the rest velocity of the H$\alpha$ emission cavity,
then the maximum blueshifted velocity we detected is closer to $\simeq$ 200 km s$^{-1}$
(see Fig.~\ref{velocity}).

Finally, these spectra indicate that the remnant's H$\alpha$ emission with
V$_{\rm rest}$ $\simeq +50$ to $-150$ km s$^{-1}$ is present across 4-5 arcsec along the
N-S slit, with faint extensions farther northward with rest frame radial
velocities of $-60$ to $-85$, $-110$, and $-135$ km s$^{-1}$. We speculate that
one of these might be associated with the narrow filament along the remnant's
northern limb seen in the direct H$\alpha$ image (Fig.~\ref{Ha_images}, right panel).


\section{Discussion}

The bright SNR S8 in the dwarf galaxy IC~1613 has been the subject
of nearly a dozen investigations since its first identification by Sandage in 1971. These studies have led to a variety of
descriptions of the S8 nebula and conclusions about its nature.  Below we
discuss its basic physical parameters derived from both our new findings and
earlier observations,
compare S8 to other young SNRs, and then state our conclusions about its nature. 

\subsection{S8's Physical Properties}
\subsubsection{Shock Velocity and Electron Density}

The remnant's expansion velocity as measured from our observed H$\alpha$
emission profile suggested a maximum velocity around 180 km s$^{-1}$.
Comparisons of its spectral line ratios with a variety shock models suggest shock velocities a bit lower, around 100 - 150  km s$^{-1}$ \citep{Peimbert1988, Lozinskaya1998}. 

In contrast, \citet{Dopita2019} claimed to only find good agreement of S8's
observed spectra using shock models having two very different shock velocities;
a fairly slow one $\sim$ 50 km s$^{-1}$ plus a much faster one
$\sim$230 km s$^{-1}$ in a ratio of 2:3. While such a mixed velocity model
predicted overly strong \ion{He}{2} and [\ion{O}{3}] line emissions compared to
observations, it did match the observed [\ion{Fe}{7}] line strength.  

A range of shock velocities would be consistent with the presence numerous [\ion{Fe}{2}] emission
lines and the presence of [\ion{O}{3}] line emission which
requires a minimum shock velocity $\sim$100 km s$^{-1}$ \citep{Raymond1979}.
Analysis of S8's optical spectra by \citet{Peimbert1988} suggested 
shock velocities as high as 160 km s$^{-1}$ to explain S8's optical spectrum,
whereas \citet{Rosado2001} suggested an even higher value V$_{\rm s}$ of
170 km s$^{-1}$.  

However, the presence of high ionization coronal lines, especially
that of [\ion{Fe}{14}], indicates much higher shock velocities
at least $\sim350$ km s$^{-1}$ \citep{DM1979}. The ratio of the strength of the
[\ion{Fe}{10}] 6374 relative to [\ion{Fe}{14}] 5303 around 1 (Table~\ref{Table_3}) suggests
a electron temperature around $1.6 \times 10^{6}$ K under the assumption of a
single temperature cause by collisional ionization and excitation
\citep{Nussbaumer1970} which requires a velocity $\sim400$ km s$^{-1}$.  

The simultaneous presence of dozens of low ionization emission lines together
with all the commonly reported optical coronal lines typically observed in gas at
temperatures $1 - 3 \times 10^{6}$ K clearly signals a wide range of
shock velocities likely generated by large variations in preshock gas densities
like that of a multi-phase ISM cloud--intercloud arrangement.  Exactly such a
scenario has been proposed in several other SNRs and actually for S8 by
\citet{Lozinskaya2009} who, not knowing about S8's coronal line emissions,
based this conclusion on the remnant's X-ray luminosity.

As discussed above, the electron density of the remnant's emission filaments from the [\ion{S}{2}]
6716/6731 line ratio indicates $n_{e}$ = $ 1000 - 1600$ cm$ ^{-1}$. 
From this we can
estimate the preshock density from the equation 
\begin{equation}
{\rm N_{[S~II]}} \simeq 31 - 45~ (\rm V_{\rm s}/100~{\rm km~s^{-1}})^2 (n_{\rm o}/{cm^{-3}}) 
\end{equation}  
where n$_{\rm o}$ is the preshock electron density and V$_{\rm s}$ is the shock
velocity responsible for the [\ion{S}{2}] emission \citep{Dopita1979,Russell1990,Rosado2001}.  
Taking N$_{\rm [S~II]}$ =  $\sim 1400$ and adopting a shock velocity range of $120 -150$ km s$^{-1}$,
we find n$_{\rm o}$ $\sim10 - 30$ cm$^{-3}$. A similar preshock density range
was found by \citet{Peimbert1988}, \citet{Lozinskaya1998}, and \citet{Rosado2001}.

\subsubsection{Mass}

We can estimate the emitting mass of S8 using the observed H$\alpha$ flux.  Our
extinction corrected H$\beta$ flux measurement of $1.3 \times 10^{-13}$ erg
cm$^{-2}$ s$^{-1}$ using a long N-S $1.4''$ wide slit corresponds to a total S8
H$\alpha$ flux value of $\simeq 6 \pm 1.5 \times 10^{-13}$ erg cm$^{-2}$
s$^{-1}$. Adopting this extinction corrected value, an electron density $N_{e}$
= $1400 \pm 200$ cm$^{-3}$, and a distance to IC 1613 of 725 kpc, we estimate
the mass of ionized hydrogen, $M_{H^{+}}$ = $85 \pm 24 M_{\odot}$ using
\begin{equation}
{\rm M_{H^{+}}}=\frac{m_p~ 4\pi d^2~ F^c({\rm H}\alpha)}{h\nu_{{\rm H}\alpha} ~ \alpha_{{\rm H}\alpha}^{eff}(H^0,T_e)N_e}
\end{equation} 
\noindent where $m_p$ is the mass of the proton, $F^c({\rm H}\alpha)$ is the
extinction corrected H$\alpha$ flux, $h\nu_{{\rm H}\alpha}$ is the energy of an
H$\alpha$ photon, and $\alpha_{{\rm H}\alpha}^{eff}(H^0,T_e)$ is the effective
recombination coefficient of H$\alpha$ equal to $8.7 \times 10^{-14}$ erg
cm$^{-2}$ s$^{-1}$ for $10^{4}$ K \citep{Osterbrock2006}.  Assuming an abundance ratio of H:He of 10:1 
by number, this leads to a total mass estimate of $119 \pm 34 
M_{\odot}$.  This is an order of magnitude less than the 1900
$M_{\odot}$ estimated by \citep{Peimbert1988} of swept of material but much more
than the 15 $M_{\odot}$ they estimated for a thin emitting shell. 

The emitting mass of S8 likely consists of both swept-up local ISM material
and any pre-SN mass loss from the progenitor.  Assuming a ellipsoid shell
dimensions of 6 pc x 6 pc x 9 pc, consistent with S8's projected angular
size of $3\farcs5 \times 5\farcs0$, and a local H~I ISM density $\sim 1 - 3$ cm$^{-3}$
\citep{Lozinskaya2001}, the remnant could have swept up $35 - 105$ 
$M_{\odot}$. The upper mass range of this estimate is consistent with the
notable increased H~I 21 cm emission in maps of the region around the S8
remnant \citep{Lozinskaya2002,Lozinskaya2009}.
   
\subsubsection{Age and SN Energy}

Assuming the S8 remnant is in an adiabatic 
expansion phase, we can use the Sedov expression 
\begin{equation} {\rm E_{o}} =
1.37 \times 10^{42}~{\rm n_{o}}~{\rm V_{s}^2} ~ {\rm R_{s}^3} 
\end{equation}
to estimate the energy of the S8 SN explosion where $E_{o}$ is the SN energy in
ergs, n$_{\rm o}$ is the preshock density per cm$^{-3}$ associated with the estimated shock
velocity, V$_{s}$ in km s$^{-1}$ and the shock radius, $R_{s}$ in pc.  Adopting
V$_{s}$ = 180 km s$^{-1}$ based on the remnant's strong blueshifted H$\alpha$ velocity 
measured from the
central cavity,
n$_{\rm o}$ $\sim$ 20 cm$^{-3}$ associated with S8's H$\alpha$ emission
as discussed above, and an average radius R$_{\rm s}$ = 7.5 pc, then E$_{\rm o}$
$\sim 4 \times 10^{50}$ erg.

We can also estimate the age of the S8 remnant 
using the Sedov expression which relates age to radius and
blast wave by the expansion timescale equation 
\begin{equation}
  t = 0.4 \times ({\rm radius/V_{blast}}).
\end{equation}  
Taking S8's average angular diameter of $\simeq4\farcs25$ which translates to a radius of $7.5$ pc at 725 kpc, and assuming
a minimum blast wave velocity of 350 km s$^{-1}$ based on the presence of [\ion{Fe}{14}] emission, 
gives a maximum age of $\sim$ 8400 yr. However, X-ray
observations suggest a much faster blast wave velocity of 660 - 1100 km s$^{-1}$ \citep{Schlegel2019} which
would imply an age of  just $\sim$ 2700 - 4400 yr.

One can check this estimated age range by using the 
estimated SN energy calculated above in the Sedov expression for remnant radius,
\begin{equation}
  r = 1.54 \times 10^{19} ~ {\rm cm} ~ {\rm E_{51}^{1/5}} ~{\rm n_{o}^{-1/5}} ~ {\rm t_{1000 yr}^{-3/5}}  
\end{equation}  
where ${\rm E_{51}}$ is SN energy in units of $10^{51}$ erg, ${\rm n_{o}}$ is the 
density in cm$^{-3}$, and ${\rm t_{1000 yr}}$ is the SNR age in units of 1000 yr.
Assuming an age of 3500 yr, a SN energy of $0.4 \times 10^{51}$ erg and a density $\sim 1$ cm$^{-3}$
consistent with H~I measurements around S8, we find a remnant radius, r $=6.8$ pc consistent with
the remnant's average radius of 7.5 pc and observed dimensions of $12 \times 18$ pc.

\begin{deluxetable*}{lccclll}[htp]
\tiny
\centerwidetable
\tablecolumns{7}
\tablecaption{Comparison of S8 with LMC Remnants N49 and N63A}
\tablewidth{0pt}
\tablehead{ \colhead{Property} &    \colhead{S8}                 &  \colhead{N49}            &  \colhead{N63A}     & \colhead{S8 Refs\tablenotemark{a}} &
\colhead{N49 Refs\tablenotemark{b}} & \colhead{N63A Refs\tablenotemark{c}}   }
\startdata
 Physical Dimensions           &  12 pc $\times$ 18 pc           &  $16 - 18$ pc             &   16 pc                & 1, 2, 3    & 1          & 1          \\
 Estimated Age                 &  $2700 - 4400$ yr               &  $3000 - 5400$ yr         & $2000 - 5000$          & 1, 4, 5    & 1, 2       & 1, 2       \\
 Estimated Nebula Mass         &  $119 \pm34 $ M$_{\odot}$       &  $207 \pm66 $ M$_{\odot}$ &  $<450$ M$_{\odot}$    & 1          & 3          & 1          \\
 Expansion Velocities          &  $+120$; $-245$ km s$^{-1}$ &$+150$; $-220$ km s$^{-1}$  &$+65$; $-245$  km s$^{-1}$ & 1, 4, 6, 7 & 4          & 3          \\
 Shock Velocities              &  $50 - 230$ km s$^{-1}$         & $50 - 270$ km s$^{-1}$    & $100 - 125$  km s$^{-1}$ & 1, 4, 7, 8, 9 & 5, 6, 7    & 4, 5       \\
 Blast Wave Velocity           &  $600 - 1100$  km s$^{-1}$      &  $\sim730$ km s$^{-1}$ & $> 350$ km s$^{-1}$ & 4, 5     & 1          & 6, 7       \\
 Filament Density              &  $1400 \pm 200$ cm$^{-3}$       & $500 - 3500$  cm$^{-3}$  &$900 - 1400$ cm$^{-3}$     & 1, 4, 6, 8 & 1, 3, 5, 7 & 6, 8       \\
 Preshock Density              &  $10 - 30$  cm$^{-3}$           &  $20 - 940$  cm$^{-3}$    & $30 - 100$  cm$^{-3}$  & 1, 4, 6, 8 & 1          & 4, 6       \\
 Coronal Line Emissions        &  [Fe VII] to [Fe XIV]           &   [Fe VI] to  [Fe XIV]    &  [Fe XI], [Fe XIV]     & 1, 9       & 5, 8, 9    & 9, 10      \\
 $[$\ion{Fe}{14}$]$ 5303/H$\beta$& 0.8                           &  1.2                      &  $0.9 - 3.0$           & 1, 9       & 5          & 9          \\
 $[$\ion{Fe}{11}$]$ 7892/H$\beta$& 1.0                           &  1.8                      &  0.3                   & 1, 9       & 5          & 10          \\
  He I  5876/H$\beta$          &    11                           &   11                      &  11                    & 1, 4, 8, 9 & 1, 10      & 8, 9, 10    \\
  He II 4686/H$\beta$          &    5                            &   5                       &  2                     & 1, 4, 8, 9 & 5, 10, 11  & 8, 9, 10    \\
 H$\alpha$ Luminosity  & $3.8 \times 10^{37}$  erg s$^{-1}$ &$2.9 \times 10^{37}$erg s$^{-1}$& $2.1 \times 10^{35}$erg s$^{-1}$ &1 & 1          & 8, 9      \\
 X-ray Luminosity      & $5.6 \times 10^{36}$  erg s$^{-1}$ &$6.4 \times 10^{36}$erg s$^{-1}$& $1.9 \times 10^{37}$erg s$^{-1}$ & 5 & 12        & 11         \\
 Radio Spectral Index, $\alpha$&   $-0.57 \pm0.054$             &   $-0.58 \pm0.04$          & $-0.74 \pm0.2$         & 4            & 13       & 12          \\
\enddata
\label{Table_4}
\tablenotetext{a}{S8 References: 1) this paper; 2) \citet{Sandage1971}; 3) \citet{DDB1980}; 4) \citet{Lozinskaya1998}; 5) \citet{Schlegel2019};
                                 6) \citet{Rosado2001}; 7) \citet{Lozinskaya2009}, 8) \citet{Peimbert1988}; 9) \citet{Dopita2019}.  }
\tablenotetext{b}{N49 References: 1) \citet{Vancura1992}; 2) \citet{Park2012};
                                  3) \citet{Melnik2013}; 4) \citet{Chu1988}; 5) \citet{Dopita2019};  6) \citet{Dennefeld1986}; 7) \citet{Pauletti2016};
                                  8) \citet{Murdin1978};  9) \citet{DM1979}; 10) \citet{Russell1990}; 11) \citet{Dopita2016}; 12) \citet{Maggi2016}; 13) \citet{Dickel1998}.   }
\tablenotetext{c}{N63A References: 1) \citet{Warren2003}; 2) \citet{Hughes1998}; 3) \citet{Chu1988}; 4) \citet{Dopita1979}; 5) \citet{Lasker1981};  6) \citet{Shull1983};  7) \citet{Rosado1986};
                                   8) \citet{Levenson1995}; 9) \citet{DM1979}; 10) \citet{Russell1990};
                                   11) \citet{Maggi2016}  12) \citet{Bozzetto2017}.   }
\end{deluxetable*}

\subsection{The Nature of the S8 Supernova Remnant}

Following Sandage's initial identification of S8 as an H II region, it 
was classified as a SNR based on optical spectra showing the characteristically
strong [\ion{S}{2}] emission relative to H$\alpha$ of shocked gas seen in SNRs
\citep{Smith1975,DD1983,Peimbert1988} plus the detection of a nonthermal radio
spectrum \cite{Dickel1985}.  However, a simple SNR classification was challenged 
when \citet{AM1985} claimed a Wolf-Rayet (WR) star was present along it
northeastern limb based on flux measurements in narrow passband filters centered
around \ion{He}{2} 4686 \AA.  A follow-up study by \citet{Massey1987} found its
spectrum `quite peculiar' and suggested the supernova remnant likely contained
a Wolf-Rayet (W-R) star based on the presence of broad \ion{He}{2} line emission
arising from a point-like source with strong continuum emission. 

Subsequent spectra by \citet{AM1991}, however, found no unusual broadening of
its \ion{He}{2} line emission leading them to discount their previous assessment
that S8 harbored a W-R star.  They concluded S8 was likely a young
SNR, but noted its spectrum was unlike that of any SNR in M31 and M33. They also 
pointed out the presence of continuum emission plus \ion{He}{2} and
[\ion{Fe}{2}] emissions resembled those seen in the Crab Nebula, while also noting
that other young SNRs exhibit strong \ion{He}{2} and [\ion{Fe}{2}] emissions
such as the LMC remnants N49, and N63A.

While \cite{Peimbert1988} estimated S8 was a fairly evolved SNR with an age
$\sim$ 20,000 yr, later studies concluded it was relatively young.
\citet{Lozinskaya1998} estimated S8's age as $\sim
3000 - 6000$ yr. They also suggested it was created by a supernova explosion inside 
an H I shell and compared it to the young LMC remnant N49. 

\citet{Rosado2001} and \citet{Schlegel2019} suggested instead it resembled more
the LMC remnant N63A.  While \citet{Dopita2019} found S8's spectral
similarities to both N49 and N63A,  both \citet{AM1991} and
\citet{Schlegel2019} discussed S8 as a possible `plerion' type of remnant like
the Crab Nebula. Recently,  \citet{Schlegel2019} concluded it was likely a
mixed morphology or composite type remnant containing both a outer thermal
emission shell enclosing nomthermal emission like that of a pulsar wind nebula.

One similarity of S8 to the Crab Nebula is that both exhibit significant continuum
emission, which in the case of the Crab is due to a bright pulsar wind nebula off its
powerful central 33 ms pulsar.  \citet{Schlegel2019} noted that S8's filled
X-ray emission structure in resolved {\sl Chandra} images are suggestive of a Crab-like
plerion or composite SNR.  However, besides obvious differences in age
and expansion velocities, S8's optical spectrum is unlike that of the Crab
Nebula, in particular it lacks the Crab's unusually strong nitrogen and helium lines
\citep{Davidson1973,Fesen1982b,MacAlpine1991,MacAlpine2008,Satterfield2012,Sibley2016}.

On the other hand, as shown in Tables~\ref{Table_1} to \ref{Table_3}, S8's
optical spectrum is remarkably similar to that of two young core-collapse SNRs
in the LMC, namely N49 and N63A. Considering IC~1613's even lower metallicity
than the LMC leading to weaker oxygen, nitrogen and sulfur emission lines, S8
exhibits an optical spectrum especially close to that observed for these young
LMC remnants.

However, whereas \citet{Rosado2001} and \citet{Schlegel2019} suggest S8 is more
like N63A than N49, we find N49 to be the better match to the S8 remnant. Both
have similar [\ion{S}{2}], [\ion{N}{2}], and [\ion{O}{3}] line strengths
relative to H$\alpha$, nearly identical electron density sensitive [\ion{S}{2}]
line ratios, similar [\ion{O}{3}] electron temperatures, relative strengths of
[\ion{Fe}{2}] emission lines and coronal Fe emissions.

In addition, besides exhibiting strikingly similar optical spectra, S8 and N49 
are of similar size, estimated age and mass, shock velocities, and
optical and X-ray luminosities as shown in Table~\ref{Table_4}.  Although the
N49 remnant is known for possessing the Soft Gamma-Ray Repeater, SGR
0526-66, no compact object has been identified in S8 based on its optical or X-ray
observations \citep{Schlegel2019}.

Finally, there is the puzzle concerning S8's strong optical continuum
emission, clearly seen in Figures~\ref{Sloan} and \ref{Ha_images} and it was this that
initially led to speculations about a continuum point source(s)
present within the remnant.  SNRs exhibiting any appreciable
continuum are exceedingly rare, with the Crab Nebula being the most famous case.
It was the presence of S8's continuum emission that caused
\citet{Massey1987} to propose that S8 might be a Crab-like remnant. 

In view of the strength of S8's continuum emission, it is somewhat
surprisingly that only Massey and collaborators emphasized the unusual finding
of strong continuum emission in S8, a property absent in the vast majority of
SNRs.  Although \citet{Lozinskaya1998} noted that some continuum emission was
indeed present in S8, they called for additional observations to determine its
extent and intensity and did not discuss possible origins except to say that it
could be consistent with a faint superimposed star in the south-central portion
of the remnant.  It is also surprisingly that neither \citet{Peimbert1988} nor
\citet{Dopita2019} make any mention of the presence of coincident continuum
emission despite it being obvious in even much lower quality spectra (e.g.,
\citealt{Massey1987}).

The solution to S8's puzzling continuum emission lies in N49's optical
properties.  Using a 6100 \AA\ continuum filter (FWHM = 130 \AA) much like our
continuum 6071 \AA \ filter (FWHM = 260 \AA) which avoids any strong emission
lines, \citet{Vancura1992} found N49 to have a surprising amount of continuum
emission plus a filamentary morphology nearly identical to that seen in the
remnant's H$\alpha$ emission.  From their H$\alpha$ and 6100 \AA \ images, they
estimated a reddening corrected continuum flux $\simeq$ 4\% that of H$\alpha$.
This flux was within 20\% of their estimated H and He recombination continuum
along with a minor contribution from two photon continua.

\begin{deluxetable*}{lcc}[htp]
\tiny
\centerwidetable
\tablecolumns{3}
\tablecaption{Possible Extragalactic SNRs Like S8}
\tablewidth{0pt}
\tablehead{ \colhead{Property} &    \colhead{M33}      &  \colhead{NGC 300}   }
\startdata
 Remnant ID                    &      L10-039                  &  S26          \\
 Physical Dimensions           &        16 pc                  &  $13 \times 15$ pc     \\
 Estimated Age                 &      \nodata                  &  $3300 \pm 700$ yr     \\
 Expansion Velocities          &   $440$ km s$^{-1}$     &$\sim400$ km s$^{-1}$   \\
 $[$\ion{S}{2}$]$ 6716/6731    &    $0.75$                     &     $0.92$             \\
 Filament Density              &   $1600$ cm$^{-3}$            &$\simeq1000$  cm$^{-3}$ \\
 Coronal Line Emissions        &  [Fe XIV]                     &    \nodata             \\
 Observed H$\alpha$ Luminosity & $9.6 \times 10^{36}$ erg s$^{-1}$  & $1.2 \times 10^{37}$  erg s$^{-1}$  \\
 X-ray Luminosity              & $3.1 \times 10^{36}$ erg s$^{-1}$  & $1.5 \times 10^{37}$  erg s$^{-1}$  \\
\enddata
\label{Table_5}
\tablerefs{M33: \citet{Long2010}, \citet{Long2018}, \citet{Blair1988}; 
                             NGC 300: \citet{Blair1997}, \citet{Millar2011}, \citet{Read2001}, \citet{Gross2019} }
\end{deluxetable*}

We find S8's continuum emission in our 6071 filter to also be $\simeq$4\% of
that of H$\alpha$ emission.  However, considering our 6071 filter's wider
bandpass, this is only about half that seen in N49. Despite this, the fact that
S8's continuum emission closely matches the size and morphology to that seen in
H$\alpha$, like that seen in N49, argues that S8's continuum  emission is also due
to H and He recombination continuum. 

S8's continuum emission may be stronger at shorter wavelengths based on its
appearance in the Sloan u band image, where its morphology does not seem
as concentrated toward the remnant's southwestern limb as in the other images
and where the remnant is brightest in H$\alpha$. If true, this raises the
possibility of a pulsar wind nebula inside the remnant displaced farther to the
east.  Such an eastern displacement is in the direction where
\citet{Schlegel2019} noted evidence, albeit weak, for a possible point source
in {\sl Chandra} X-ray images of the S8 remnant. 

\subsection{Are There Other Extragalactic SNRs Like S8?} 

The finding that S8 is a young ``N49-like'' SNR raises the question how unusual
are these young and luminous remnants, and might there other similar
extragalactic SNRs already detected but not identified as such?  Currently,
there are over 1500 extragalactic SNRs detected in a dozen or so nearby galaxies
\citep{Long2017,Long2019}. A key element that sets S8 and N49 apart from most other
extragalactic SNRs is their relative youth  meaning they still possess high
blast wave velocities ($\sim 500 - 1000$ km s$^{-1}$). If these remnants expand
into an extensive multi-phase ISM cloud, this will lead to a range of
shock velocities, high filament densities $>10^{3-4}$ cm$^{-3}$, high optical
and X-ray luminosities, H and He recombination continuum, and coronal line
emissions from lower density regions. 

More than two dozen optical extragalactic SNR searches have shown that the
majority of remnants are old and large with diameters greater than 20 pc. Few remnants
show a density sensitive [\ion{S}{2}] 6716/6731 line ratio below unity,
indicative of $n_{\rm e}$ $\geq 1000$ cm$^{-3}$
\citep{Long2010,Leonidaki2013,Blair2014b,Long2018}.  This suggests that N49-like
remnants might be identifiable using the criteria of a small physical size
(dia.\ $\leq$ 20 pc) plus [\ion{S}{2}] 6716/6731 ratios $< 1$. In addition, if such
remnants had fast shocks impacting dense CSM or ISM clouds, they would also
exhibit high optical and X-ray luminosities like seen for S8 and N49.  

A non-exhaustive review of extragalactic SNR surveys reveals a few potential S8
and N49 like remnants (see Table~\ref{Table_5}). For example, out
of 197 M33 SNRs identified \citep{Long2010,Long2018}, the
remnant L10-039 (DDB-7, \citealt{DDB1980}) shows the highest H$\alpha$
surface brightest, the largest H$\alpha$ flux, the greatest H$\alpha$ velocity
width (FWHM = 443 km s$^{-1}$; \citealt{Blair1988}) and is one of only two remnants with a
6716/6731 ratio $<$1.0.  Moreover, a plot of its optical spectrum shown in
\citet[][their Fig.\ 2]{Long2018} appears to show the presence of 
[\ion{Fe}{14}] 5303 \AA \ emission which would support a shock velocity above
350 km s$^{-1}$.  

The S26 remnant in NGC 300 also shows some S8-like properties. It is optically
the brightest SNR in this spiral galaxy and the only one with a [\ion{S}{2}]
derived density of $\sim1000$ cm$^{-3}$ \citep{Blair1997}. It is also one of the
smallest SNRs in the galaxy (dia.\ $\simeq$ 15 pc), appears relatively young
with an estimated age $\sim3300$ yr and a shock velocity $\sim450$ km s$^{-1}$
based on its X-ray properties \citep{Gross2019}.    

\subsection{The Missing SN Ejecta in Young Remnants}

Lastly, neither of the suspected core-collapse remnants, S8 and N49, show any
signs of emissions from high-velocity metal-rich ejecta as one might expect due
to their relatively young ages ($\leq 5000$ yr). This is in contrast to
high-velocity, O-rich ejecta knots and filaments present in some other
similarly young remnants. 

For example, the $3700- 4500$ yr old Galactic remnant Puppis A shows several
1500 km s$^{-1}$ O-rich ejecta knots and filaments. The much younger $1000 -
2000$ yr old SMC remnant 1E0102-7219 consists almost entirely of O-rich ejecta
expanding at $-2500$ to +3500 km s$^{-1}$
\citep{Tuohy1983,Finkelstein2006,Vogt2010}, roughly half that of the younger
350 yr old Galactic SNR, Cassiopeia A. 

In an {\sl HST} search of young SNRs in M83, none of the 50+ smallest and
presumably young remnants (dia.\ $<$ 15 pc) showed any optical emissions
indicating the presence of O-rich ejecta, leading
\citet{Blair2014a,Blair2014b} to wonder about the cause of the missing
ejecta-dominated remnants in such small remnants.  Follow-up Gemini-S GMOS
spectra confirmed the lack of obvious ejecta-enhanced abundances and revealed
no high expansion velocities.  Their sample of small remnants included ones as
small as Cas A (dia.\ = 5 pc) but exhibited only narrow, low-velocity line
emissions indicative of ordinary radiative ISM shocks. A few objects such as
B12-150 exhibited a [\ion{S}{2}] 6716/6731 ratio of 0.75 indicating an electron
density around 1600 cm$^{-3}$, much like S8 and N49 which might have suggested
a young age. But no high-velocity emissions were detected.  

In light of these findings, objects like S8 and N49 may represent a later
evolutionary phase of younger remnants like the small SNRs seen in M83 which
lack of optical emission from high-velocity ejecta.  For there to be detectable
late-time ejecta emission, there must be enough dense ejecta clumps to be
readily visible, plus a significant source of excitation and ionization.
Possible sources include strong reverse-shock heating, high-velocity
interaction with surrounding CSM or ISM, and X-ray and UV emissions from a
pulsar-wind nebula. 

Regardless of the energy source, whether or not SN ejecta are optically bright
is determined by the object's emitting column density and whether the electron
density and temperature of the gas are within a certain range of values
($n_{\rm e} \geq$ 10 cm$^{-3}$, T$_{\rm e}$  $\sim 10^{4} - 10^{5}$ K).  We
speculate that the underlying cause for the lack of optically detectable ejecta
in some $2500
- 5000$ yr old SNRs may be that only a small fraction of the remnant's mass of
  high-velocity, O-rich ejecta are presently dense enough and at the right
temperature to be visible optically. It may be that in many thousand year old
remnants the density of the majority of their ejecta clumps and filaments are
below detection limits due to knot disruption after passage of the reverse
shock and by knot ablation and fragmentation, like that seen in Cas A's ejecta
knots \citep{Fesen2011}, during the centuries long interaction phase with the local
ISM.

\section{Conclusions}

In this paper we have presented high resolution optical images and spectra of
the compact SNR, S8, in the nearby dwarf galaxy IC~1613. H$\alpha$
images of the remnant show a sharply defined crescent shaped nebula $3\farcs5
\times 5\farcs0$ in size which is 12 pc x 18 pc at IC~1613's distance of 725 kpc.
Narrow passband images clear of the remnant's bright emission lines reveal a
coincident and unexpectedly bright continuum nebulosity exhibiting a size and
morphology like that seen for the remnant's line emissions.  

Low-dispersion spectra reveal numerous low-ionization line emissions such as
[\ion{O}{1}] and [\ion{Fe}{2}], along with higher-ionization emission lines
including \ion{He}{2} and optical coronal lines ([\ion{Fe}{7}], [\ion{Fe}{10}],
[\ion{Fe}{11}], and [\ion{Fe}{14}]) indicating a wide range of shock velocities
present, from $\sim$ 50 to over 350 km s$^{-1}$, due to multi-phase ISM with
preshock densities of $\sim1 - 30$ cm$^{-3}$.  Higher resolution optical
spectra indicate an expansion velocity around 180 km s$^{-1}$ with a central
cavity $\simeq45$ km s$^{-1}$ wide. From reddened correct H$\alpha$ flux, we
estimate a total nebula mass of $119 \pm 34$ M$_{\odot}$ and a remnant age of
$\simeq2700 - 4400$ yr. 

Combining past research plus our new data, we conclude the S8 supernova remnant
is remarkably like those seen in the similarly young LMC remnant N49 including
physical size, shock and expansion velocities, filament densities, optical line
strengths, X-ray and optical luminosities, along with optical coronal lines and
continuum emissions. Both remnants are relatively young possessing a high
velocity blast wave ($600
- 1500$ km s$^{-1}$) which has interacted with dense interstellar clouds; in
the case of N49  molecular clouds, and H~I clouds for S8. 

Further study of the S8 remnant could include much higher spatial resolution optical
imaging using {\sl HST} to investigate the remnant's line emission structure in
comparison with N49's extensive and thin filamentary appearance. Deep narrow
passband, line emission-free images could also explore the remnant's continuum
emission structure in relation to that seen in its line emission, and investigate 
continuum emission coincident with a possible point X-ray source hinted at in {\sl
Chandra} images.

Because of S8's unusual properties, it would also be of interest to search and
identify other similar young SNRs.  Follow-up optical spectra of the remnants
L10-039 in M33, S26 in NGC 300 and other similar objects could explore their
nature through spectral line modeling, expansion velocity measurements, and the
possible detection of high-velocity ejecta.  Assembling a larger set of young
$2500 - 5000$ yr old SNRs would lead to a better understanding of the general
properties of remnants during their early phases of expansion and evolution.

\acknowledgements

We thank the referee for helpful comments and suggestions, and Justin Rupert
and  Eric Galayda and the MDM staff for outstanding assistance that made these
observations possible.  We also thank Taniana Lozinskays, Roger Chevalier, Dan
Milisavljevic, and Mike Beitenholz for helpful comments.  K.E.W.\ acknowledges
support from Dartmouth's Guarini School of Graduate and Advanced Studies, and
the Chandra X-ray Center under CXC grant GO7-18050X.  This work is part of
R.A.F's Archangel III Research Program at Dartmouth.  Finally, we would like to
acknowledge Mike Dopita not only for his research on S8 and young LMC SNRs but
his enormous legacy spanning more than four decades on the study and spectral
modeling of supernova remnants and other emission line nebulae.

\facilities{Hiltner (OSMOS), McGraw (CCDS)}

\software{PYRAF \citep{pyrafcite}, Astropy v4.0 \citep{AstropyCiteA,AstropyCiteB}, 
ds9 \citep{ds9cite}, L.A.\ Cosmic \citep{vanDokkum2001}, OSMOS Pipeline (thorosmos: \url{https://github.com/jrthorstensen/thorosmos})}

\bibliography{ref.bib}

\end{document}